# MarcoPolo-R narrow angle camera: a three-mirror anastigmat design proposal with a smart finite conjugates refocusing optical system

Jacopo Antichi[a], Massimiliano Tordi[b], Demetrio Magrin[c], Roberto Ragazzoni[c], Gabriele Cremonese[c]

[a]Osservatorio Astrofisico di Arcetri, Largo Enrico Fermi 5, 50125 Firenze, Italia;
[b]Space Technologies s.r.l., Viale Combattenti A. d'Europa 9/G, 45100 Borsea, Italia;
[c]Osservatorio Astronomico di Padova, Via Vicolo Osservatorio 5, 35100 Padova, Italia.

## ABSTRACT

MarcoPolo-R is a medium-class space mission proposed for the 2015-2025 ESA Cosmic Vision Program with primary goal to return to Earth an unaltered sample from a primitive near-Earth asteroid (NEA). Among the proposed instruments on board, its narrow-angle camera (NAC) should be able to image the candidate object with spatial resolution of 3 mm per pixel at 200 m from its surface. The camera should also be able to support the lander descent operations by imaging the target from several distances in order to locate a suitable place for the landing. Hence a refocusing system is requested to accomplish this task, extending its imaging capabilities. Here we present a three-mirror anastigmat (TMA) common-axis optical design, providing high-quality imaging performances by selecting as entrance pupil the system aperture stop and exploiting the motion of a single mirror inside the instrument to allow the wide image refocusing requested, from infinity up to 200 m above the NEA surface. Such proposal matches with the NAC technical specifications and can be easily implemented with present day technology.

## 1. INTRODUCTION

The main goal of the MarcoPolo-R mission is to return samples from a primitive asteroid. Presently, the mission target is 175706 1996 FG3, a pristine binary NEA that exhibits water deep near-infrared (NIR) absorption bands onto the reflected light spectrum. Its water content has been considered essential to shed light on to the prebiotic chemistry start-up on Earth. Indeed, some groups of carbonaceous chondrites accreted with a significant organic and water content have been discovered yet, but meteorites reaching the Earth surface are biased towards the ones with the highest strength. How many the water participated at the catalysis of organic compounds on Earth is then a question still open. MarcoPolo-R will assess the abundance of carbon and water present in selected NEA, answering important questions on the amount and complexity of organics and water delivered to the early Earth. These topics have been considered extremely relevant for a dedicated mission by the Planetary Science Community in Europe to merit the participation to the ESA 2015-2025 Cosmic Vision selection Program.

When dealing with a returned sample from space, several characterization requirements are necessary: a global characterization that means the measure of the properties of the NEA on a global scale, a local characterization that means to indentify a potential sampling site and finally a context characterization that means making measures at the selected site onto the NEA surface. To this aim, the MarcoPolo-R will mount several imaging instruments for different characterization levels. Among them the scope of the NAC is to provide high quality narrow angle images in the whole VISible up to the NIR wavelength range, by which the sampling site will be definitely selected among a sample of 5 [1].

## 2. INSTRUMENT CONCEPT

High spatial resolution images is of paramount importance for the overall target body characterization in term of: surface topography and distribution of morphological features (e.g. boulders, craters, fractures), generation of digital model of the surface, analysis of the fragmentation and accretion evolutions, bulk composition of the body (size, shape, rotational properties), overall characteristics like orbit, rotation, size, mass, gravity and density.

NAC imaging may also be required for complementary details of the shape model obtained through a wide-angle camera (WAC) mounted on board of MarcoPolo-R and for a close characterization of any putative satellite body that might be discovered around the main target by the WAC.

Primitive asteroids are intrinsically dark objects. This is why the assumed average albedo for the target is equal to 0.06. In this illumination regime high-contrast images can be obtained only with a camera optics, whose modulus transfer function (MTF) is limited only by diffraction and without central obstruction at the pupil plane, which in turn, for a given image space focal ratio, degrades the best diffraction limited MTF shape profile. This is why axial symmetric catoptric designs should be discarded as their corresponding dioptric ones; these lasts for the bigger spacing requested that, in this case, scales linearly with focal length of the system. Hence, the NAC optical design is based upon the off-axis TMA configuration and follows the heritage of the OSIRIS-NAC on board of the Rosetta mission, which is in-flight and working according to the original specifications.

TMA are commonly adopted to correct for all the Seidel aberrations with high accuracy at the cost of realizing off-axis and globally aspheric optical systems. This option is optically and mechanically manageable when the beam size is modest as in this science case, where the entrance pupil is fixed to be lower than 100 mm. In this frame, the option to select as entrance pupil the system aperture stop revealed itself to be useful to easier the stray light baffling [2]. With this option in, the TMA layout returns to be an off-axis pupil and off-axis focal plane design, slightly different from the on-axis pupil design proper to the OSIRIS-NAC [3].

## 3. INSTRUMENT SPECIFICATIONS AND LAYOUT DEFINITION

The main instrument specifications of the MarcoPolo NAC are related to spatial resolution and to the MTF quality at the Nyquist spatial frequency. In detail, it should be able to image an object at 200 m distance with spatial resolution equal to 3mm/pixel. In fact, it is foreseen that the main payload will enter in orbit with the target NEA at 5 Km and will move close in up to 200 m above the surface. On the other side, the request on MTF quality is given at the Nyquist frequency imposed by the pixel size of the adopted detector: a 2048 x 2048 hybrid Si_PIN array with 10 μm square pixels [4]. Hence, the NAC Nyquist spatial frequency returns to be 50 mm$^{-1}$. The entrance pupil size has been defined to be 82.5 mm in order to have sufficient flux from the surface of the object with assumed average albedo equal to 0.06 and assumed maximum exposure time equal to 0.01 second [4].

Table 1: MarcoPolo-R NAC main technical specifications.

| | | |
|---|---|---|
| Pupil diameter | mm | 82.5 |
| Focal ratio | | 8 |
| Field of view (FOV) | deg x deg | 1.7 x 1.7 |
| Pixel scale at 200 m | mm/px | 3 |
| MTF at the Nyquist spatial frequency | % | > 60% over the entire FOV |
| Spectral range | nm | 400-950 |
| Detector pixels number | | 2048 x 2048 |
| Pixel size | μm x μm | 10 x 10 |

To avoid movable mechanical components for the requested series of broad, medium and narrow band chromatic filters, such filters are conceived to be contiguous and spaced strips deposited onto the detector window. In this configuration, the NAC will be able to image simultaneously up to 8 contiguous portions of the target at different wavelengths [2]. To avoid the detector moves during the requested refocusing of the camera, from 5 Km up to 200 m from the NEA surface, the selected option is the one to allow the 2$^{nd}$ TMA mirror to be free of a piston movement and of a de-centering movement with respect to its axis of symmetry. This option is very different from the one selected for OSIRIS-NAC on board of Rosetta, which consisted in the exploitation of the optical path differences of the chromatic filters mounted onto a mechanical wheel to guide refocusing and chromatic filters selection in one shot [2]. That option actually foresees no movements at all for the instrument optics, at the coast of a chromatic dependent refocusing procedure that strongly limits the exploitable mission concept. A part from this aspect, the major drawback of this kind of optical option is the injection of unwanted ghost light due by the double reflection within the chromatic filters substrate and finally imaged onto the focal plane [5]. On the contrary, the MarcoPolo-R NAC refocusing option is ghost-free, a part from the detector window substrate impact, at the cost of an active optics with trivial movements: piston and de-centering.

Finally, the best condition setup for a safety alignment is established once the axes of symmetry of the 3 mirrors are coincident i.e. once the TMA is in a *common axis* layout. In this proposed layout, M2 returns to be tilted with respect to the common axis of symmetry of M1 and M3. However, the common axis condition is verified once the pivotal point of M2 is taken into account instead of the mirror vertex point. The TMA common axis is obtained then by considering the line passing through the vertex points of M1, M3 and the pivotal point of M2, which in turn has been defined to be the mirror geometrical center.

## 4. OPTICAL DESIGN WITH A SMART REFOCUSING PROPOSAL

The optical design is a common axis TMA, optimized to compliant with the specifications listed in Table 1. The aperture stop is the entrance window of the system: in this way the entrance pupil is fixed to be 82.5 mm, independently onto the refocusing configuration, from 5 Km to 200 m object distance position, that foresees a piston movement and a de-centering movement of M2 while letting the absolute positions of M1 and M3 fixed. This is the most important characteristic of our layout and it marks the distinction of this TMA solution with respect to the OSIRIS-NAC one, where the aperture stop is fixed to be M2 [3]. Hence in that case no movements at all are allowed without varying the entrance pupil size. On the contrary, in this proposal we let the entrance window to be the system entrance pupil. This option, already chosen in a previous study [2], returns to be perfectly compliant with our refocusing option, which in turn is based upon two independent movements on M2.

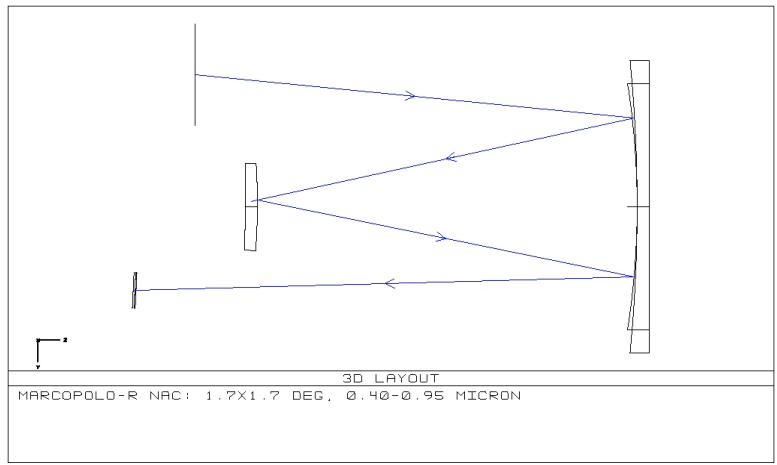

Figure 1: NAC layout proposal: the length from the entrance window to M1 & M3 is 360 mm, from M2 to M1 & M3 is 307.27 mm, from M2 to the detector surface is 105 mm and from M1 & M3 to the detector surface is 412.27 mm.

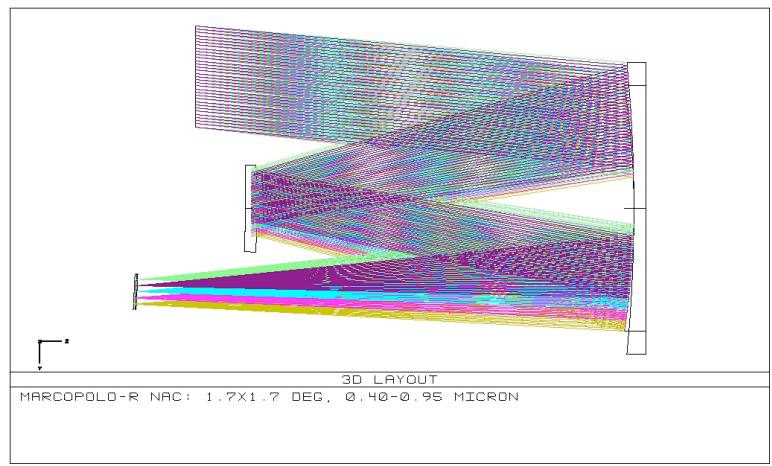

Figure 2: NAC layout proposal: imaging is optimized to diffraction limit within 1.7x1.7 deg$^2$ and from 400 up to 950 nm.

According to definitions [3], the selected layout belongs to the family of off-axis pupil TMA designs: the on-axis chief ray is not pinned to the TMA common axis, and to the off-axis FOV TMA designs: the on-axis chief ray is not coaxial to the entrance pupil axis of symmetry, see Figure 1. Moreover, the selected layout has been optimized looking at solutions with at least 1 spherical mirror, as suggested by literature on the TMA optics alignment, e.g. [6]. To this aim we applied the same choice made for the OSIRIS-NAC design i.e. we let M3 to be spherical [3].

Table 2: MarcoPolo-R NAC mirrors optical parameters.

| Optical element | M1 | M2 | M3 |
| --- | --- | --- | --- |
| Vertex curvature radius | 1243.93 mm | -512.20 mm | 657.39 mm |
| Conic constant | -2.19 | -1.32 | 0.00 |

Table 3: MarcoPolo-R NAC mirrors geometrical parameters in the case the image plane conjugated to infinite distance.

| Optical element | M1 | M2 | M3 |
| --- | --- | --- | --- |
| Relative distance | 307.27 mm | 307.27 mm | 105.00 mm |
| Offset | 0.000 mm | 0.000 mm | 0.000 mm |
| Tilt | 0.00 deg | 0.36 deg | 0.00 deg |

Table 4: MarcoPolo-R NAC mirrors geometrical parameters in the case the image plane conjugated to 5 Km distance.

| Optical element | M1 | M2 | M3 |
| --- | --- | --- | --- |
| Relative distance | 307.35 mm | 307.35 mm | 105.00 mm |
| Offset | 0.000 mm | 0.011 mm | 0.000 mm |
| Tilt | 0.00 deg | 0.36 deg | 0.00 deg |

Table 5: MarcoPolo-R NAC mirrors geometrical parameters in the case the image plane conjugated to 4 Km distance.

| Optical element | M1 | M2 | M3 |
| --- | --- | --- | --- |
| Relative distance | 307.37 mm | 307. 37 mm | 105.00 mm |
| Offset | 0.000 mm | 0.013 mm | 0.000 mm |
| Tilt | 0.00 deg | 0.36 deg | 0.00 deg |

Table 6: MarcoPolo-R NAC mirrors geometrical parameters in the case the image plane conjugated to 3 Km distance.

| Optical element | M1 | M2 | M3 |
| --- | --- | --- | --- |
| Relative distance | 307.40 mm | 307.40 mm | 105.00 mm |
| Offset | 0.000 mm | 0.018 mm | 0.000 mm |
| Tilt | 0.00 deg | 0.36 deg | 0.00 deg |

Table 7: MarcoPolo-R NAC mirrors geometrical parameters in the case the image plane conjugated to 2 Km distance.

| Optical element | M1 | M2 | M3 |
| --- | --- | --- | --- |
| Relative distance | 307.46 mm | 307.46 mm | 105.00 mm |
| Offset | 0.000 mm | 0.027 mm | 0.000 mm |
| Tilt | 0.00 deg | 0.36 deg | 0.00 deg |

Table 8: MarcoPolo-R NAC mirrors geometrical parameters in the case the image plane conjugated to 1 Km distance.

| Optical element | M1 | M2 | M3 |
| --- | --- | --- | --- |
| Relative distance | 307.65 mm | 307.65 mm | 105.00 mm |
| Offset | 0.000 mm | 0.053 mm | 0.000 mm |
| Tilt | 0.00 deg | 0.36 deg | 0.00 deg |

Table 9: MarcoPolo-R NAC mirrors geometrical parameters in the case the image plane conjugated to 800 m distance.

| Optical element | M1 | M2 | M3 |
|---|---|---|---|
| Relative distance | 307.75 mm | 307.75 mm | 105.00 mm |
| Offset | 0.000 mm | 0.066 mm | 0.000 mm |
| Tilt | 0.00 deg | 0.36 deg | 0.00 deg |

Table 10: MarcoPolo-R NAC mirrors geometrical parameters in the case the image plane conjugated to 600 m distance.

| Optical element | M1 | M2 | M3 |
|---|---|---|---|
| Relative distance | 307.90 mm | 307.90 mm | 105.00 mm |
| Offset | 0.000 mm | 0.088 mm | 0.000 mm |
| Tilt | 0.00 deg | 0.36 deg | 0.00 deg |

Table 11: MarcoPolo-R NAC mirrors geometrical parameters in the case the image plane conjugated to 400 m distance.

| Optical element | M1 | M2 | M3 |
|---|---|---|---|
| Relative distance | 308.23 mm | 308.23 mm | 105.00 mm |
| Offset | 0.000 mm | 0.133 mm | 0.000 mm |
| Tilt | 0.00 deg | 0.36 deg | 0.00 deg |

Table 12: MarcoPolo-R NAC mirrors geometrical parameters in the case the image plane conjugated to 200 m distance.

| Optical element | M1 | M2 | M3 |
|---|---|---|---|
| Relative distance | 309.20 mm | 309.20 mm | 105.00 mm |
| Offset | 0.000 mm | 0.268 mm | 0.000 mm |
| Tilt | 0.00 deg | 0.36 deg | 0.00 deg |

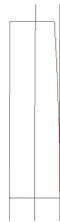

Figure 3: The design foresees a common tilt of M2 equals 0.36 deg: this angle is highlighted by the (red) tangent to the mirror vertex and by a line passing onto the vertex too, but parallel to the rotation angle at the mirror geometrical center.

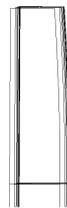

Figure 4: The design foresees a piston (relative distance variation between M1 and M2) movement and a de-centering movement (offset) with respect to the common axis of symmetry. The reference position is the TMA common axis configuration that corresponds to the optical setup where the image plane is conjugated to an infinite distance.

# 5. OPTICAL PERFORMANCES

Table 13: Spot and average MTF diagrams: spots are well within 2 pixels (Airy disk at 600 nm is shown for comparison) and the MTF value at the Nyquist frequency is beyond 60% wherever in the 1.7 x 1.7 deg$^2$ FOV. (a) Image plane conjugated to infinite, (b) to 5 Km distance, (c) to 4 Km distance, (d) to 3 Km distance, (e) to 2 Km distance.

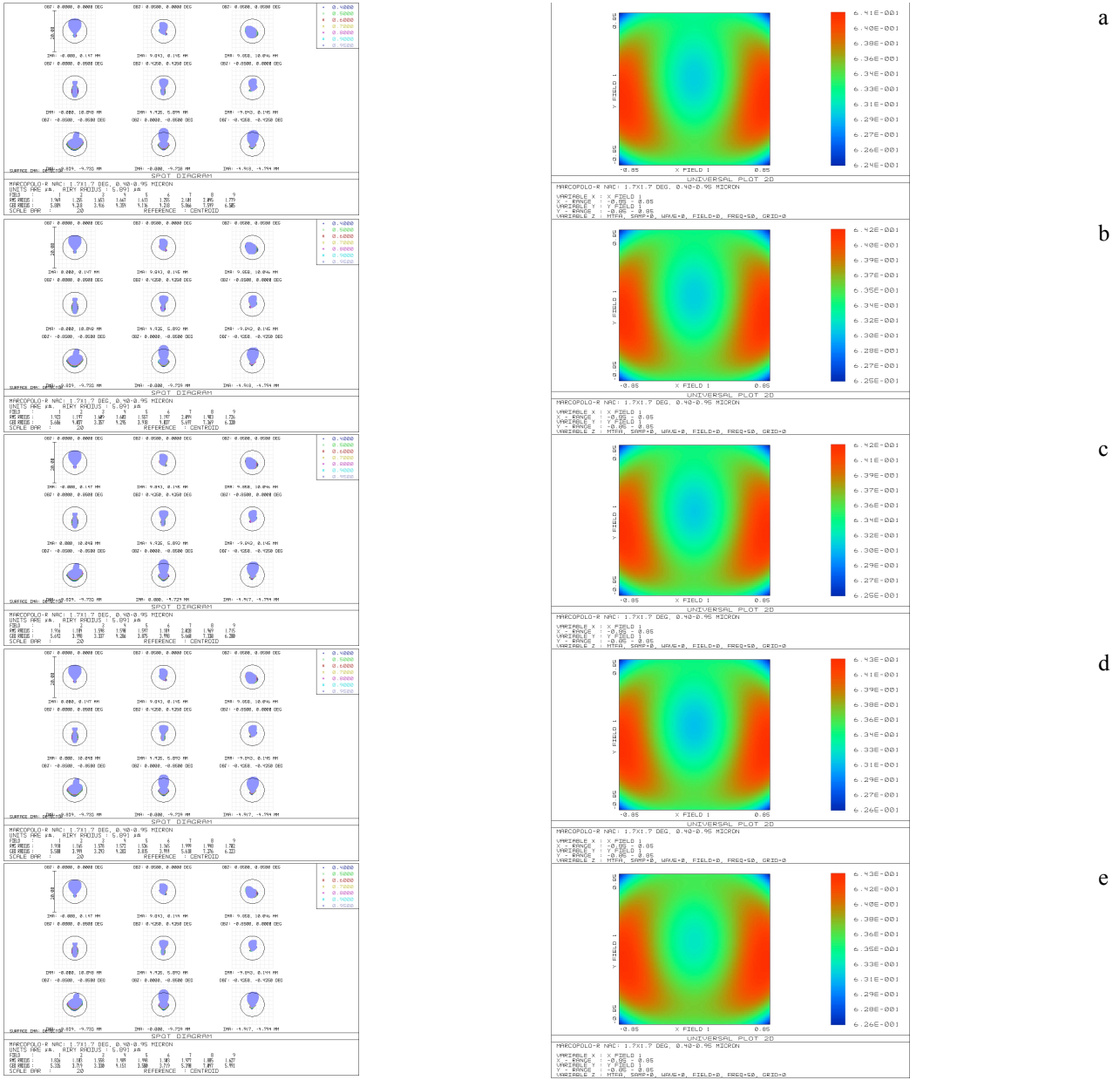

The performances of this proposal layout meet the specifications of the MarcoPolo-R NAC: diffraction limit is achieved at 600 nm and average MTF is well beyond 60% at the Nyquist frequency equals 50 mm$^{-1}$ when the system is allowed to re-focus throughout the M2 movements (piston and de-centering), ranging from the configuration where the image plane is conjugated to infinite up to 2 Km distance.

Table 14: Spot and average MTF diagrams: spots are well within 2 pixels (Airy disk at 600 nm is shown for comparison) and the MTF value at the Nyquist frequency is beyond 60% wherever in the 1.7x1.7 deg$^2$ FOV. (f) Image plane conjugated to 1 Km distance, (g) to 800 m distance, (h) to 600 m distance, (i) to 400 m distance, (l) to 200 m distance.

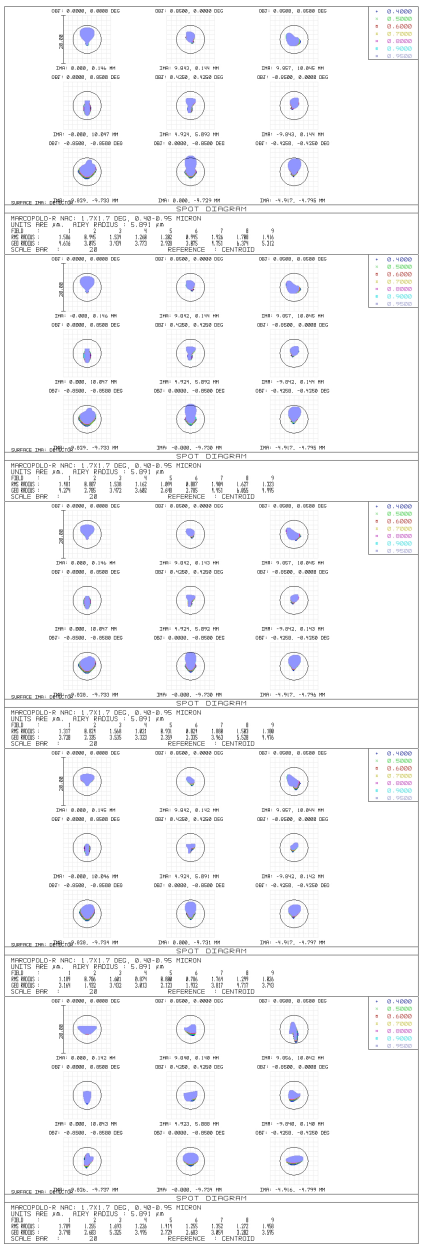
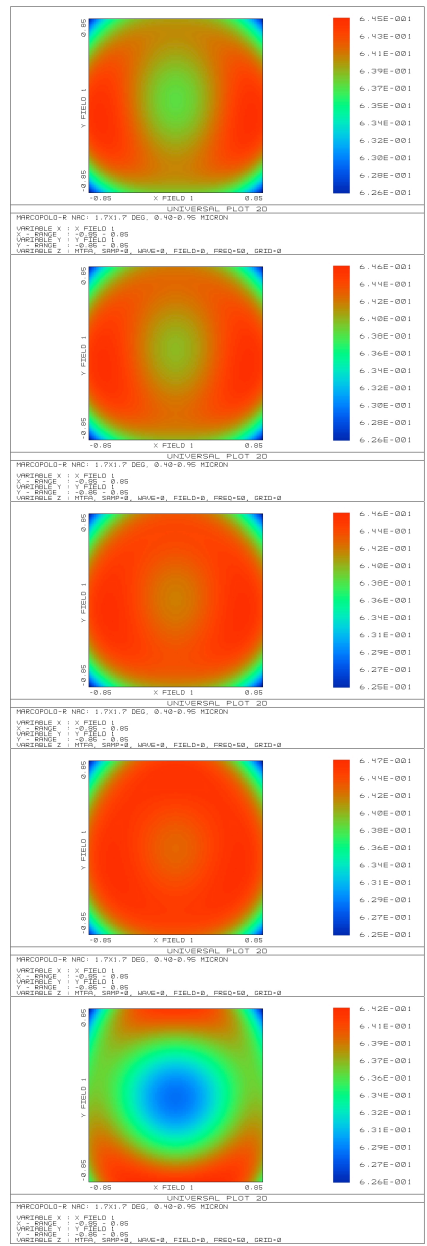

f

g

h

i

l

The performances of this proposal layout meet the specifications of the MarcoPolo-R NAC: diffraction limit is achieved at 600 nm and average MTF is well beyond 60% at the Nyquist frequency equals 50 mm$^{-1}$ when the system is allowed to re-focus throughout the M2 movements (piston and de-centering), ranging from the configuration where the image plane is conjugated to 1Km up to 200 m distance.

# 6. SMART REFOCUSING SENSITIVITY ANALYSIS

A sensitivity analysis has been performed to fix the required positioning accuracy of the M2 translation stage. With this aim we pursued a tolerance analysis of the NAC layout without letting the system compensating the final focal position to accomplish the fixed criterion, which in turn was selected to be the MTF over the entire FOV, evaluated at the Nyquist frequency. A maximum sensitivity of the MTF of 1% was tolerated. The resulting positioning errors of M2 when the system undergoes a re-focus are listed in the following Table.

Table 15: Positioning accuracy of the M2 translation stage.

| Piston | Decentering |
|---|---|
| ± 0.01 mm | ± 0.005 mm |

These 2 numbers are essential both to fix the tolerances of the translation stage mechanics and to evaluate the loss of performances in case of a refocusing system failure. Within the positioning accuracy, the translation stage results to be a linear guide, see Figure 5, with refocusing distances listed in Table 16. The guide is tilted with respect to the optical axis by 7.94 deg. With the positioning accuracies of Table 15, it follows that it shall be integrated into the system with an available mechanical tolerance ranging between 7.55 deg and 8.33 deg.

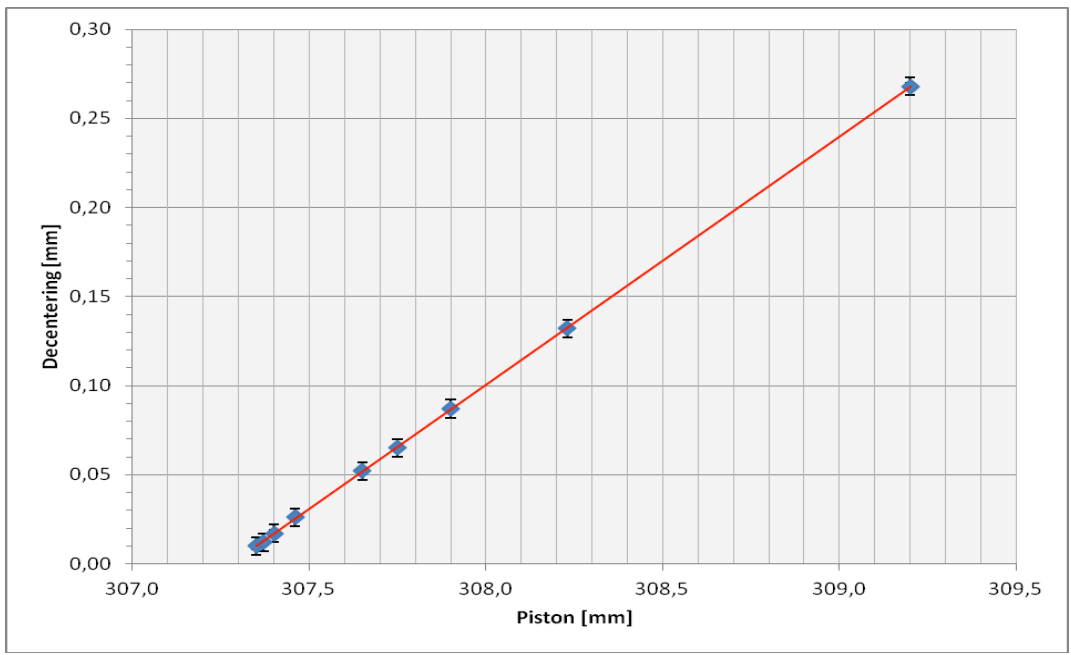

Figure 5: The M2 translation stage results a linear combination of piston and decentering, error bars refer to Table 15.

Table 16: Piston and decentering vs. object distance.

| Object distance [Km] | Piston [mm] | Decentering [mm] |
|---|---|---|
| 5.0 | 307.35 | 0.011 |
| 4.0 | 307.37 | 0.013 |
| 3.0 | 307.40 | 0.018 |
| 2.0 | 307.46 | 0.027 |
| 1.0 | 307.65 | 0.053 |
| 0.8 | 307.75 | 0.066 |
| 0.6 | 307.90 | 0.088 |
| 0.4 | 308.23 | 0.133 |
| 0.2 | 309.20 | 0.269 |

The sensitivity analysis can also be used to obtain the positioning error of the actuator, providing the movement of the mirror, along its linear guide. With this aim, we considered the positioning error as a quadratic sum of the accuracy and the bidirectional repeatability. Using a parametric representation of the actuator movement, the resulting total admissible positioning error along the guide results to be ± 0.005 mm. We notice that it comes out from the analysis of the worst scenario: i.e. the object position conjugated closer to the camera at 200 m.

The total admissible positioning accuracy is important also to look at our smart refocusing system as a single point of failure for the whole mission. In this frame, it is mandatory to evaluate the loss of performance that might be expected in case of a failure of the actuator. As when a failure occurs the conjugation position of the mirror is not predictable, we examined every conjugation scenario between infinity up to the closest distance. Results are shown in Figure 6 and they have been obtained as follows: first, the conjugating position of the mirror has been fixed a priori, as if the failure of the actuator was occurred at that position; then we modified the object distance from the infinite up to 200 m, with 100 steps each times and we calculated at each step the resulting MTF at the Nyquist frequency. As expected, the closer the conjugation position is the quicker the MTF drops: for instance, if the actuator was blocked at a conjugate position of 1 Km, the MTF would fall below 50% as an object distance smaller than about 790 m or larger than about 1200 m; if the actuator was blocked at a conjugate position of 3 Km, then the MTF would fall below 50% as an object distance smaller than about 1655 m, while it would stay above 50% for every object position larger than 3 Km. Finally, at very small distances from the object (< 400 m), the NAC imaging quality is acceptable only within a very small range of distances (tens of meters) around the conjugate position.

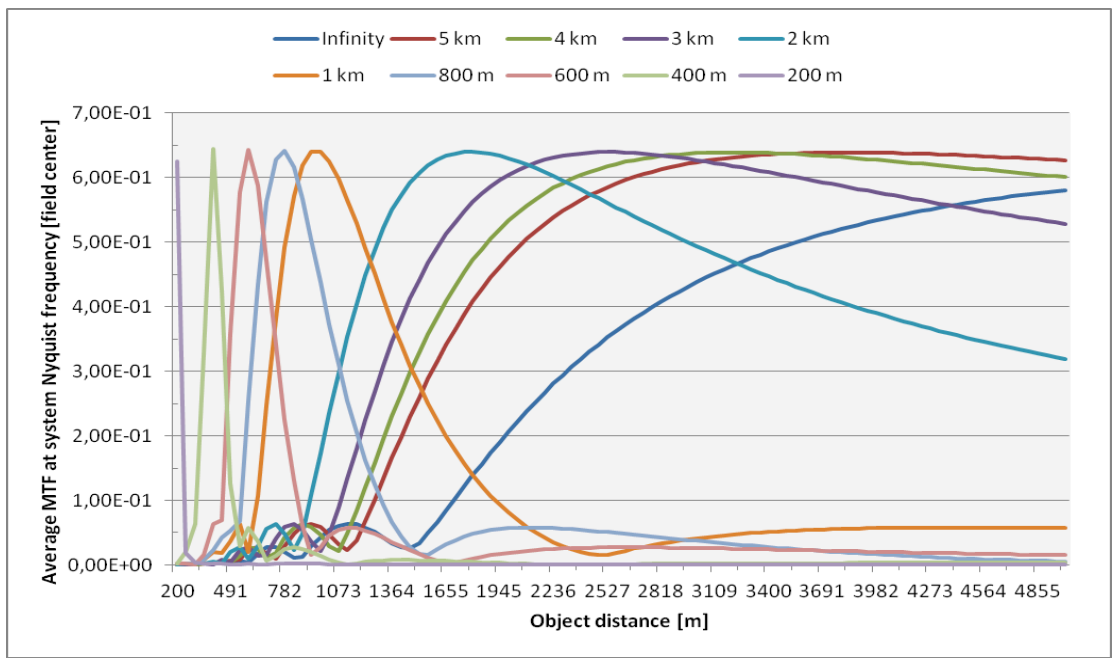

Figure 6: MTF performance at the Nyquist frequency in the case of a failure of the M2 actuator at conjugate positions ranging from infinity up to 200 m above the NEA surface.

## 7. CONCLUSIONS

We presented a TMA proposal for the narrow angle camera of the MarcoPolo-R mission, fixing as entrance pupil of the system its aperture stop placed before the primary mirror. At the current status of development, the mission concept requires that the camera is free to refocus from 5 Km to 200 m above the NEA surface. This is the strongest requirement of this optics, exceeding the same requirement of it precursor ORISIR-NAC reference mission, implementing a refocusing for conjugated distances in the range 500.000 - 1 Km [3], above the 67P/Churyumov-Gerasimenko comet.

After the evaluation of different alternatives, the best option we found consists in a pure translation of the TMA secondary mirror. As in this layout option this procedure does not impact onto the entrance pupil size, it leaves the system focal ratio unaltered. To estimate the system feasibility, we evaluated the positioning accuracy required to the actuator behind the mirror. First, we determined the movement requires a linear guide, making its implementation quite affordable and with loose mechanical tolerances. Nonetheless, the actuator implementing the active refocusing along the guide would represent a single point of failure of the whole MarcoPolo-R mission, a price paid by every opto-mechanical solution implementing moving parts on space borne instrumentation. Hence, we provided means to evaluate the loss of performances of the proposed refocusing system, in the case a failure of the actuator occurs wherever along its guide.